\documentclass[a4paper,12pt,final]{elsarticle}

\newif\ifcolor
\colortrue

\newif\ifarxiv
\arxivtrue

\ifarxiv
  \newcommand{\figuresdir}{}
  
\else
  \newcommand{\figuresdir}{figures/}
  
\fi

\usepackage[cdot]{SIunits}
\usepackage[dvips]{epsfig}
\usepackage{upgreek}
\usepackage{subeqn}
\usepackage{subfig}
\usepackage{multirow}

\renewcommand{\electronvolt}{\ensuremath{\mathrm{e\hspace{-1pt}V}}}

\newcommand{\eV}{\electronvolt}

\newcommand{\keV}{\kilo\eV}

\newcommand{\TeV}{\tera\eV}

\newcommand{\lumUnit}{\centi\meter\rpsquared\usk\reciprocal\second}

\newcommand{\Vcmcmmg}{\volt\, \centi\meter\squared\per\milli\gram}
\newcommand{\electron}{\ensuremath{\mathrm{e}}}

\newcommand{\eminus}{\ensuremath{\electron^{-}}}

\newcommand{\muon}{\ensuremath{\upmu}}
\newcommand{\muplus}{\ensuremath{\muon^{+}}}
\newcommand{\muminus}{\ensuremath{\muon^{-}}}

\newcommand{\proton}{\ensuremath{\mathrm{p}}}

\newcommand{\muonium}{\ensuremath{\mathrm{Mu}}}
\newcommand{\mucq}[1]{\ensuremath{\muon^{(#1)}}}
\newcommand{\mucplus}{\mucq{+}}
\newcommand{\muczero}{\mucq{0}}
\newcommand{\mucminus}{\mucq{-}}

\newcommand{\parfig}{figure}

\newcommand{\textfig}{figure}

\newcommand{\textFig}{Figure}
\newcommand{\texttab}{table}
\newcommand{\textTab}{Table}
\newcommand{\textsec}{section}

\newcommand{\texteqn}{equation}

\newcommand{\differential}[1]{\ensuremath{\mathrm{d}#1}}

\newcommand{\diff}[2]{\ensuremath{\frac{\differential{#1}}{\differential{#2}}}}
\newcommand{\Idiff}[2]{\ensuremath{\differential{#1}/\differential{#2}}}

\newcommand{\initialism}[1]{\textsc{#1}}

\newcommand{\PSI}{\initialism{psi}}

\newcommand{\IdTds}{\Idiff{T}{s}}

\newcommand{\Istoppingpower}{\ensuremath{(1/\rho)\,\IdTds}}
\newcommand{\sub}[2]{\ensuremath{#1_{\mathrm{#2}}}}
\newcommand{\Teq}{\sub{T}{eq}}
\newcommand{\TeqPrime}{\ensuremath{\Teq^\prime}}

\begin{document}
\begin{frontmatter}

\title{Frictional cooling of positively charged particles}
\author[address:mpi]{Daniel Greenwald\footnote{Corresponding author, deg@mpp.mpg.de}}
\author[address:mpi]{Allen Caldwell}

\address[address:mpi]{Max Planck Institute for Physics, Munich, Germany}

\begin{abstract}
  One of the focuses of research and development towards the
  construction of a muon collider is muon beam preparation. Simulation
  of frictional cooling shows that it can achieve the desired
  emittance reduction to produce high-luminosity muon beams. We show
  that for positively charged particles, charge exchange interactions
  necessitate significant changes to schemes previously developed for
  negatively charged particles. We also demonstrate that foil-based
  schemes are not viable for positive particles.
\end{abstract}


\end{frontmatter}

\thispagestyle{empty}

\section{Introduction}

One of the focuses of research and development towards the
construction of a muon collider is muon beam creation and preparation.
The short lifetime of the muon necessitates preparation of a muon beam
on time scales shorter than a microsecond. To achieve luminosities on
the order of \unit{\power{10}{34}}{\lumUnit} as in the schemes
of~\cite{alsharo19a:ez,Ankenbrandt:2007zz,Abramowicz:2004}, requires a
reduction of beam emittance---known as beam cooling---by six orders
of magnitude.

Simulations of frictional cooling~\cite{Muhlbauer:1993sb} show that it
can achieve the desired emittance reduction and potentially produce
high-luminosity muon beams~\cite{Abramowicz:2004}. Previous work on
frictional cooling, including both simulation and
experiment~\cite{Muh96}, focused on cooling \muminus\ beams and
assumed the cooling mechanism would be identical for \muplus\ beams.
However, positively charged particles, unlike negatively charged ones,
participate in charge exchange processes, and this greatly alters
their cooling. In this paper, we present calculations and simulations
of the effects of charge exchange processes on the frictional cooling
of positive particles. We also show that previous simulations of
foil-based schemes are invalid for positive particles, and rule out
the viability of such schemes.

\section{Frictional Cooling}
\label{sec:frictional_cooling}

\begin{figure*}[t]
  \centering
  \ifcolor
    \includegraphics[width=\textwidth]{\figuresdir 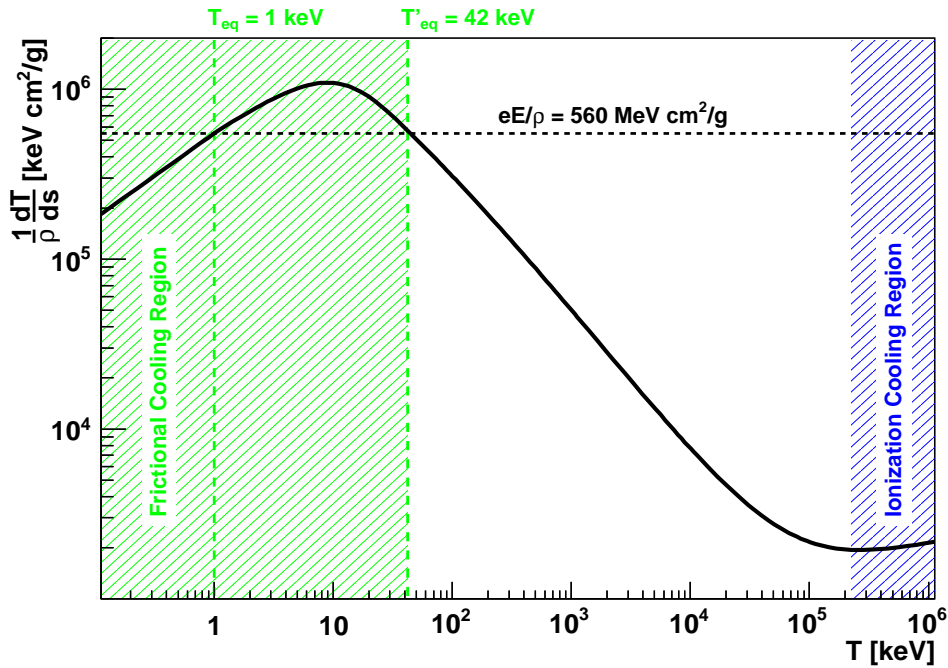}
  \else
    \includegraphics[width=\columnwidth]{\figuresdir dEdx_bw.eps}
  \fi
  \caption{Stopping power of helium on \muplus\ (velocity scaled
    from~\cite{icru49:1994tf}) and the acceleration power of an
    electric field of fixed strength $E$ for a particle of constant
    unit charge as functions of particle energy $T$.\label{fig:dEdx}}
\end{figure*}

Frictional cooling reduces the energy spread and divergence of a beam
by balancing retarding forces from interactions with a medium, with
accelerating forces from an electric field to bring the beam to an
equilibrium energy. \textFig~\ref{fig:dEdx} shows the stopping power
$S=\Istoppingpower$, where \IdTds\ is the energy loss per unit path
length, and $\rho$ is the medium density, for helium on \muplus\ and
the acceleration power ($E/\rho$) of an electric field of fixed
strength $E$ on a particle of constant unit charge. The stopping power
is velocity scaled\footnote{Except where noted, data for \muplus\
  interactions in this paper are velocity scaled from proton
  data.}~\cite{bransden:1989} for \muplus\ from the proton data given
in~\cite{icru49:1994tf}. The stopping power on \muminus\ has a similar
shape, though it is smaller in magnitude at energies below
approximately \unit{100}{\keV}~\cite{PhysRevLett.74.371,
  springerlink:10.1007/s100530170162}. When the accelerating power is
larger than the stopping power, the particle is accelerated. When the
reverse is true, the particle is decelerated. At the kinetic energies
where the two powers are equal, particles are at an energy
equilibrium. If the stopping power is greater than the accelerating
power at energies above this equilibrium point and vice versa below
it, then the point is stable and attractive.

For a particle of constant charge, two equilibrium points are created:
a stable one (\Teq\ in \parfig~\ref{fig:dEdx}) at an energy below that
at which the stopping power peaks, and an unstable one at an energy
above (\TeqPrime). Particles with $T<\TeqPrime$ will be brought to
\Teq, defining the frictional cooling energy region.

\begin{table*}[t]
  \centering
  \begin{tabular}{l*{3}{r@{.}l}l@{}r}
    \hline\hline

    \Teq &
    2&0 & 2&5 & 3&0 &
    \keV\\

    \hline\hline

    \Idiff{S}{T} &
    7&55 & 6&46 & 5&69 &
    \centi\meter\squared\per\milli\gram &

    \multirow{3}{*}{
      $\left.
        \begin{array}{c}
          \\
          \\
          \\
          \\
        \end{array}
      \right\}
      \muminus \text{ in H}_{2}
      $}\\

    \cline{1-8}

    Total &
    0&60 & 0&69 & 0&83 &
     \keV \\

    Scattering &
    0&57 & 0&64 & 0&76 &
    \keV\\

    Straggling &
    0&18 & 0&27 & 0&34 &
    \keV \\
        
    \hline\hline

    Charge Exchange &
    0&07 & 0&11 & 0&17 &
    \keV &
    \muplus \text{ in He}\\

    \hline\hline
  \end{tabular}
  \caption{Energy spreads of frictionally cooled beams of \muminus\ in 
    H$_2$ gas from~\cite{Muhlbauer:1993sb} and \muplus\ in He; and
    gradient of the stopping power of H$_2$ on \muminus\ from the
    parameterization in~\cite{PhysRevLett.74.371}\label{tab:muehlbauer}}
\end{table*}

Straggling of energy losses to the medium prevent a beam from becoming
truly mono-energetic in a frictional cooling scheme, inducing a spread
of the beam energy around \Teq. A study of the frictional cooling of
\muminus\ in~\cite{Muhlbauer:1993sb} found that the spread of the
energy distribution of a cooled beam is independent of the beam's
initial spread and, as \texttab~\ref{tab:muehlbauer} shows, decreases
with increasing gradient of the stopping power. In the table, the
gradient is calculated from the parameterization of the stopping power
for hydrogen on \muminus\ from \cite{PhysRevLett.74.371}, using the
measured parameters given in the paper.

Such effects are shared by ionization cooling~\cite{Neu83,
  Parkhomchuk:1983ua}, which operates at higher kinetic energies.
However, in contrast to ionization cooling, frictional cooling, which
operates at low kinetic energies, is greatly affected by nuclear
scattering, which contributes to the spread in kinetic energies of the
cooled beam to a larger extent than the straggling of energy losses.
This is due to particles scattering away from the electric field
direction, possibly even in directions opposed to that of the field;
they are then slowed down and reaccelerated to the equilibrium energy
in the direction of the field.

\section{Low-Energy Stopping Processes}

At high kinetic energies, a projectile slows down in a medium through
excitation and ionization of the medium atoms. One can neglect the
interactions with the nuclei of the medium and assume that a
positively charged projectile is stripped of all its
electrons~\cite{fano1963,ziegler2008}. Thus the stopping power of the
medium for positive particles is the same as for negative ones. This
is the energy region of ionization cooling, and the energy loss here
is comparatively simpler to model than energy loss at low energies.

At low energies, projectiles slow down by Coloumbic interactions with
both the nuclei and the electrons of the medium; furthermore the
interactions of positively charged projectiles involve more than
excitation and ionization, leading to a difference in the stopping
powers for positive and negative projectiles.

Scattering of the projectile particle off of a nucleus results in a
loss of energy and a change of direction. Though, as described above,
nuclear scattering has the largest impact on the final energy spread
of a cooled \muminus\ beam, it is not the main mechanism of energy
loss. The nuclear stopping power is orders of magnitude smaller than
the electronic stopping power for all but the lowest kinetic
energies~\cite{icru49:1994tf}.

\subsection{Electronic Stopping}
\label{sec:electronic_stopping}

The interactions of the projectile\footnote{The equations to follow
  are equally valid with \muplus\ and \muonium\ replaced by \proton\
  and H.} with the electrons of the stopping medium are the dominant
mechanisms of energy loss. In the frictional-cooling energy region
these interactions are the excitation and ionization of the medium
atoms (X),
\begin{subequations}
  \begin{eqnarray}
    \mucplus + \mathrm{X}      &\rightarrow& \mucplus + \mathrm{X}^* \label{eqn:excitation}\\
    \mucplus + \mathrm{X}^{q} &\rightarrow& \mucplus + \mathrm{X}^{q+n} + n\eminus \label{eqn:ionization}
  \end{eqnarray}
  and the capture and loss of an electron by the projectile,
  \begin{eqnarray}
    \mucplus + \mathrm{X}^q &\rightarrow& \muczero + \mathrm{X}^{q+1} \\
    \muczero + \mathrm{X}^q &\rightarrow& \mucplus + \mathrm{X}^q + \eminus \\
    \muczero + \mathrm{X}^q &\rightarrow& \mucplus + \mathrm{X}^{q-1} \label{eqn:electronloss},
  \end{eqnarray}
  where we have introduced the notation \mucq{q}\ to represent charge
  states of \muplus\ as an ion of muonium~(\muonium):
  \begin{equation}
    \begin{array}{lclcl}
      \mucplus & = & \muonium^+ & = & \muplus, \nonumber\\
      \muczero & = & \muonium   & = & \muplus\eminus,\ \mathrm{and} \nonumber\\
      \mucminus& = & \muonium^- & = & \muplus\eminus\eminus. \nonumber
    \end{array}
  \end{equation}

  It is obvious that stopping of the neutral charge state plays a role
  in the stopping of the projectile. We must also consider processes
  (\ref{eqn:excitation})--(\ref{eqn:electronloss}) with the
  replacement of \mucplus\ by \muczero, and \muczero\ by \mucminus;
  and processes involving the negative charge state:
  (\ref{eqn:excitation}) and (\ref{eqn:ionization}) with the
  replacement of \mucplus\ by \mucminus, and double electron capture
  and loss
  \begin{eqnarray}
    \mucplus + \mathrm{X}^q  &\rightarrow& \mucminus + \mathrm{X}^{q+2} \label{eqn:doublecapture} \\
    \mucminus + \mathrm{X}^q &\rightarrow& \mucplus + \mathrm{X}^{q-2} \label{eqn:doubleloss}.
  \end{eqnarray}
\end{subequations}

We write the cross sections for processes
(\ref{eqn:excitation})--(\ref{eqn:doubleloss}) for \mucplus, \muczero,
and \mucminus\ with the notation $\sigma_{qq'}(T)$, denoting the total
cross section for the interactions taking a muon of charge state $q$
and energy $T$ to a muon of charge state $q'$ (accompanied by an
energy loss). To be clear, $q$ and $q'$ refer to the charge state of
the muonium ion $(+,0,-)$, not to the charge of the muon itself, which
remains positive in these purely electromagnetic interactions.

The total stopping power of the projectile is the combination of the
individual stopping powers of the different charge states,
\begin{eqnarray}
  S(T) &=& \sum_q f^q(T)S^q(T) \nonumber\\
  &=&f^+(T)S^+(T) + f^0(T)S^0(T) + f^-(T)S^-(T), \label{eqn:stopping_power_sum}
\end{eqnarray}
where the $f^q$ are the equilibrium charge state
fractions~\cite{RevModPhys.30.1137}, which are the solutions to
\begin{equation}
  \diff{\!f^q}{x} \propto \sum_{q'}\left(f^{q'}\sigma_{q'q} - f^{q}\sigma_{qq'}\right) = 0, \qquad\forall q,
\end{equation}
and
\begin{equation}
  \sum_qf^q = 1.
\end{equation}
Since the \Idiff{\!f^{q}}{x} are taken at a fixed $T$, the $f^{q}$ are
functions of $T$. For the three-state system $\{\mucplus, \muczero,
\mucminus\}$, the equilibrium charge states are
\begin{subequations}
  \begin{equation}
    f^{q} =  A^{q}\left/\ \sum_{q'}A^{q'} \right.,
  \end{equation}
  where
  \begin{equation}
    A^{q} \equiv \sum_{i\ne j\ne q}\sigma_{ij}\sigma_{jq} + \prod_{i\ne q}\sigma_{iq}
  \end{equation}
\end{subequations}

By the nature of its measurement~\cite{icru49:1994tf}, the stopping
power shown in \textfig~\ref{fig:dEdx} is the total stopping power;
that is, the left-hand side of
\texteqn~(\ref{eqn:stopping_power_sum}).

\section{Effective Charge}

It is clear from \textsec~\ref{sec:electronic_stopping} that when
traveling through matter, \muplus\ spends some time as \muonium\ and
$\muonium^-$. This changing of charge state will have a significant
impact on the frictional cooling process, which involves the
restoration of energy losses by an electric field according to $qE$.

\begin{figure*}[t]
  \centering
  \includegraphics[width=\columnwidth]{\figuresdir 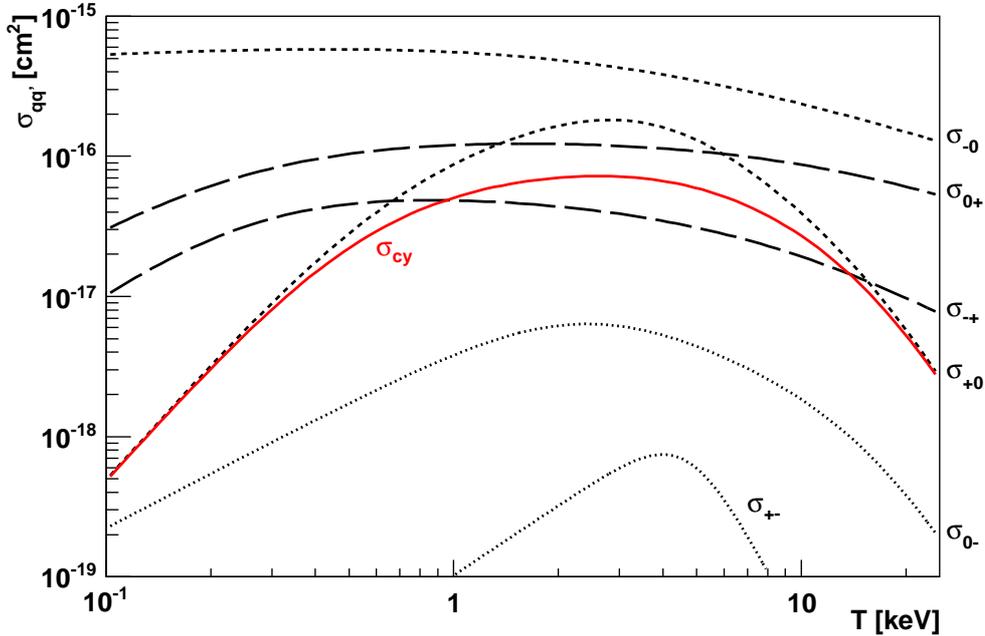}
  \caption{Charge exchange cross sections for \muplus\ in helium,
    velocity scaled from proton cross sections in~\cite{nak87}
    resulting in final states with \mucplus\ (large dashes), \muczero\
    (small dashes), and \mucminus\ (dots). The two-state
    charge-change-cycle cross section is shown in solid\ifcolor \
    red\fi.\label{fig:charge_exchange_xs}}
\end{figure*}

Figure~\ref{fig:charge_exchange_xs} shows the cross sections for the
charge exchange processes of \muplus\ in helium obtained by velocity
scaling those for protons in~\cite{nak87}. The cross sections
resulting in \mucminus\ charge states are orders of magnitude smaller
than those taking \mucminus\ to \mucplus\ or \muczero. So \muplus\
traveling in helium (and in all the materials we will discuss) is
nearly in a two-state system~$\{\mucplus, \muczero\}$.  This
simplifies the calculation of the equilibrium charge state fractions
to
\begin{equation}
  f^{+} = \frac{\sigma_{0+}}{\sigma_{+0}+\sigma_{0+}}\quad\mathrm{and}\quad
  f^{0} = \frac{\sigma_{+0}}{\sigma_{+0}+\sigma_{0+}}.
\end{equation}

\begin{figure*}[t]
  \centering
  \subfloat[Helium]{\includegraphics[width=.48\columnwidth]{\figuresdir 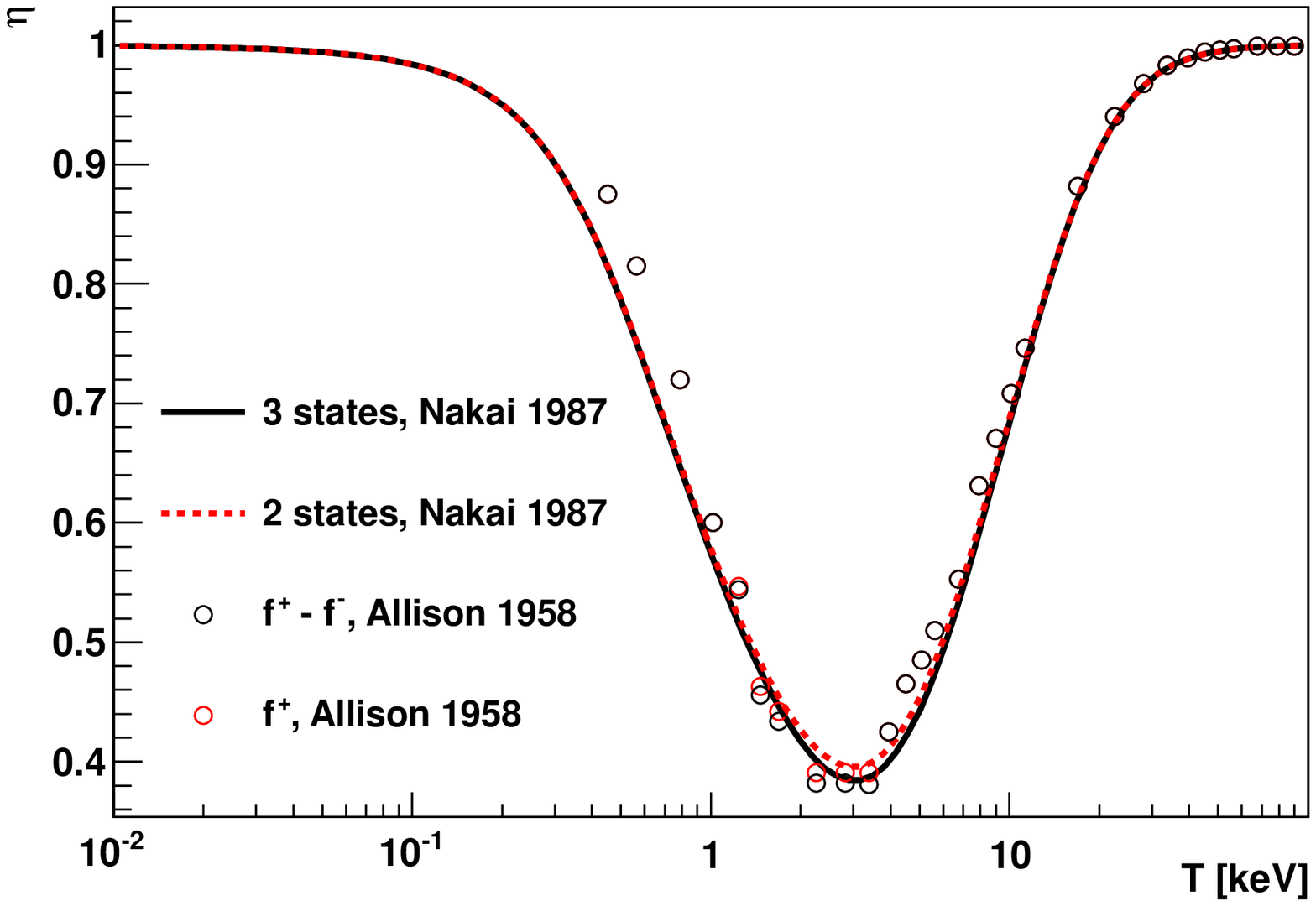}\label{fig:effective_charge_He}}
  \subfloat[Hydrogen]{\includegraphics[width=.48\columnwidth]{\figuresdir 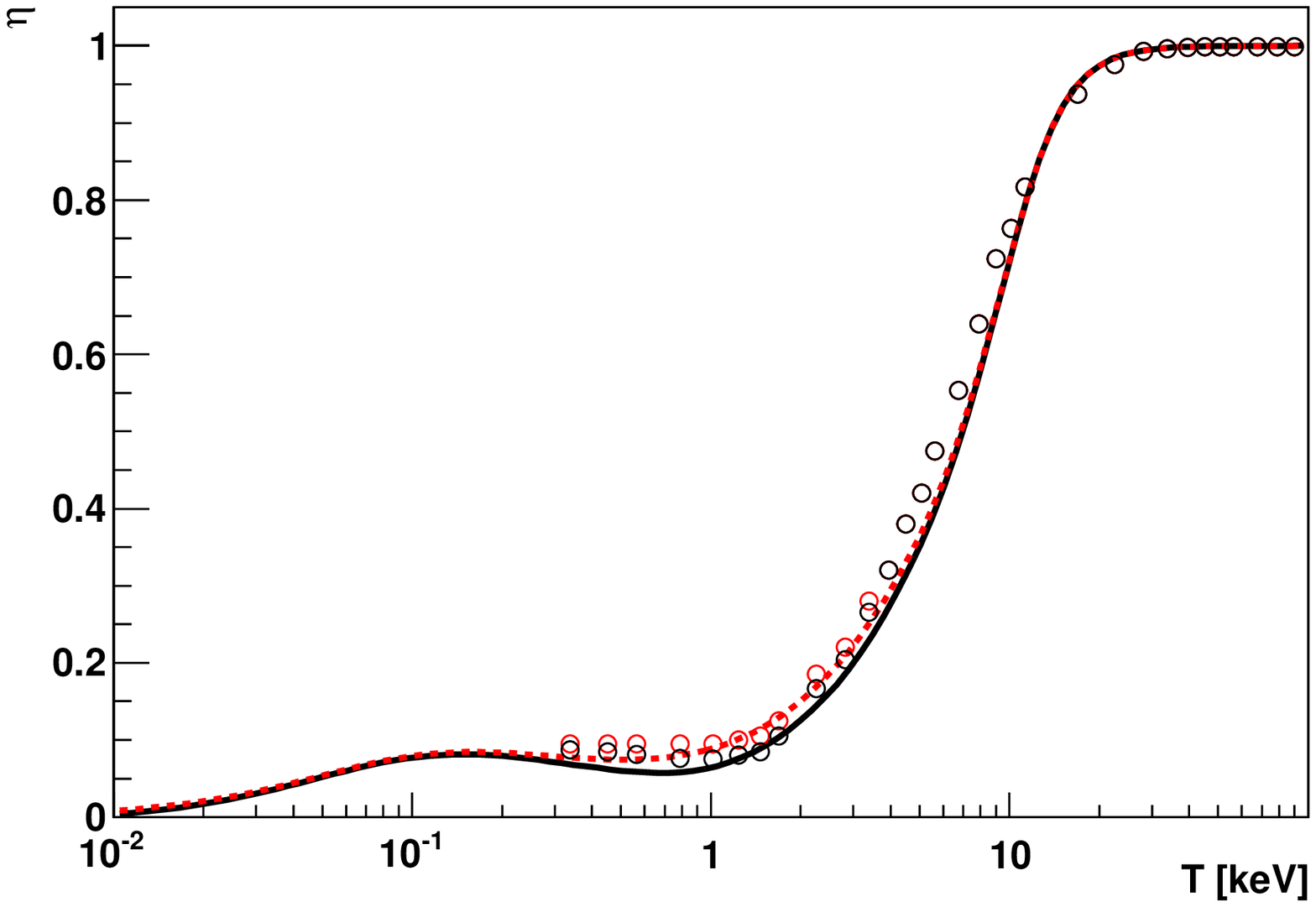}\label{fig:effective_charge_H}}\\
  \subfloat[Neon]{\includegraphics[width=.48\columnwidth]{\figuresdir 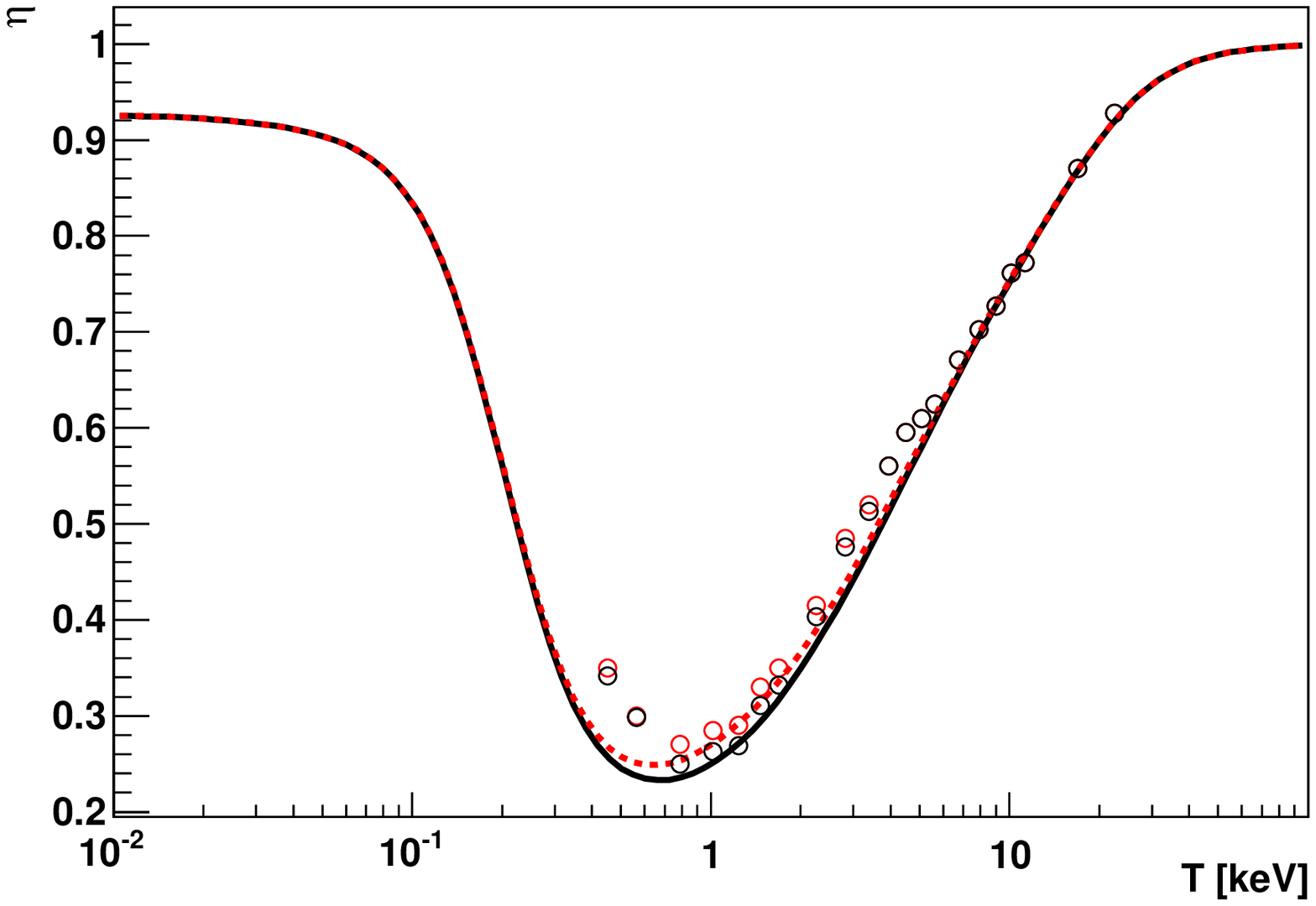}\label{fig:effective_charge_Ne}}
  \subfloat[Carbon]{\includegraphics[width=.48\columnwidth]{\figuresdir 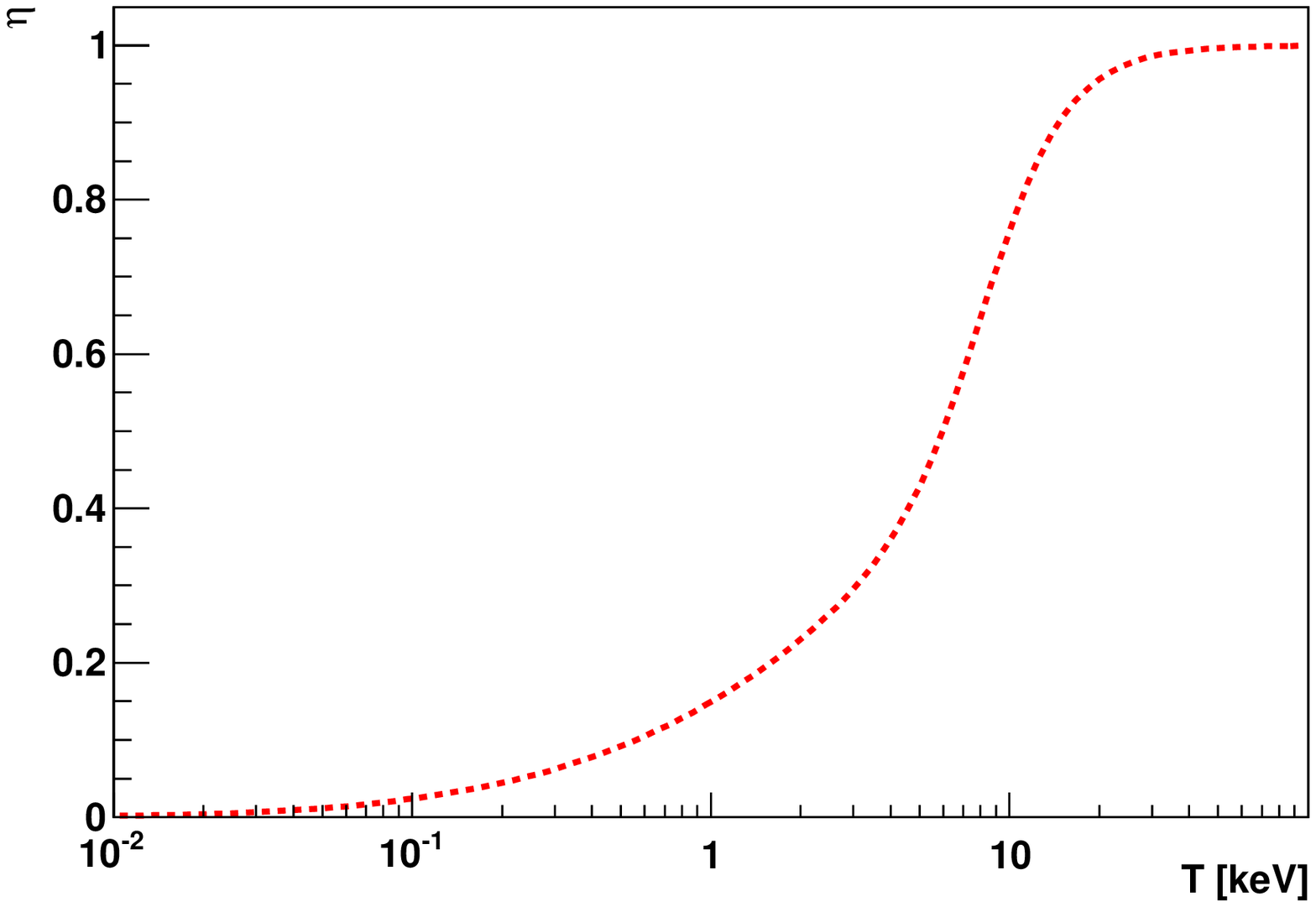}\label{fig:effective_charge_C}}
  \caption{Effective charge (line) for \muplus\ in He, H, Ne, and C
    for the three-state (solid\ifcolor\ black\fi) and two-state
    (dashed\ifcolor\ red\fi) systems. The data points show the
    equilibrium charge state fractions taken from
    \cite{RevModPhys.30.1137}\label{fig:effective_charge}.}
\end{figure*}

The mean free path for one charge change cycle
($\mucplus\rightarrow\muczero\rightarrow\mucplus$) can be calculated
from the mean free paths for the individual charge exchange processes,
$\sub{\lambda}{cy} = \lambda_{0+} + \lambda_{+0}$. This yields a cross
section for a charge exchange cycle to take place
\begin{equation}
  \sub{\sigma}{cy} = \frac{\sigma_{+0}\,\sigma_{0+}}{\sigma_{+0}+\sigma_{0+}},
\end{equation}
which we show in \textfig~\ref{fig:charge_exchange_xs}. Since the
charge exchange interactions take place over distances much shorter
than the overall distance the muon travels in the frictional cooling
scheme, we can approximate the charge of the \muplus\ by an effective
charge according to
\begin{equation}
  \eta \equiv \sum_{q}qf^{q}. \label{eqn:effective_charge}
\end{equation}
\textFig~\ref{fig:effective_charge} shows the effective charges of
\muplus\ in helium, hydrogen, and neon for both the two-state
(\sub{\eta}{\,2}) and three-state (\sub{\eta}{\,3}) systems calculated
from the charge-exchange cross sections using empirical formulae
from~\cite{Gre71} fit to measured values~\cite{nak87}. We also show
\sub{\eta}{2} for carbon; \sub{\eta}{3} could not be calculated since
the cross sections involving the negative charge state are unknown.

For comparison to the calculated effective charge, we show
experimentally measured charge state fractions
from~\cite{RevModPhys.30.1137} for helium, hydrogen, and neon:
$f^{+}$, which is the same as \sub{\eta}{\,2}; and $(f^+-f^-)$, which
is the same as \sub{\eta}{\,3}.  The calculated effective charge
matches very well with the measured data. As well, we see that
\sub{\eta}{\,2} and \sub{\eta}{\,3} differ only minutely and only over
a small range of energies. The negative charge state fraction
contributes to the three-state effective charge at percent level and
lower in all three gases.

It is important to note that helium and neon are the only materials in
which the effective charge tends to a value of or near unity at low
energies. In hydrogen and carbon (as well as water, oxygen, and
nitrogen, and therefore air) the effective charge approaches zero at
low energies.

\section{Accelerating Power \& Cooling Medium}

The effective charge of a positively charged projectile in the
retarding medium of a frictional cooling scheme can be absorbed into
the accelerating power of the electric field. In effect, this makes
the accelerating power dependent on the projectile's kinetic energy.

We reformulate here the requirements for frictional cooling stated in
\textsec~\ref{sec:frictional_cooling}: An equilibrium energy must be
established by balancing energy loss with energy gain,
\begin{equation}
  S(\Teq) = \frac{E}{\rho}\eta(\Teq).
\end{equation}
At energies above \Teq, the stopping power must be greater than the
accelerating power
\begin{subequations}
  \begin{equation}
    S(\Teq+\epsilon) - \frac{E}{\rho}\eta(\Teq+\epsilon) > 0.
  \end{equation}
  And at energies below \Teq\ the accelerating power must be greater
  than the stopping power
  \begin{equation}
    \frac{E}{\rho}\eta(\Teq-\epsilon) - S(\Teq-\epsilon) > 0.
  \end{equation}
  The last two requirements can be combined to one statement about the
  slope of the stopping power relative to that of the accelerating
  power:
  \begin{equation}
    \left.\diff{S}{T}\right|_{\Teq} - \left.\frac{E}{\rho}\diff{\eta}{T}\right|_{\Teq} \equiv S'(\Teq) - \frac{E}{\rho}\eta'(\Teq) > 0.
  \end{equation}
\end{subequations}
All three requirements can be met when
\begin{equation}
  S' - S\frac{\eta'}{\eta} > 0. \label{eqn:cooling_requirement}
\end{equation}


\begin{table*}[t]
  \centering
 
  \begin{tabular}{@{}l*{10}{r@{.}l}@{\ }r@{}}
    \hline
    \hline

    ~&
    \multicolumn{2}{l}{He} &
    \multicolumn{2}{l}{H$_2$} &
    \multicolumn{2}{l}{Ne} &
    \multicolumn{2}{l}{C} &
    \multicolumn{2}{l}{N$_2$} &
    \multicolumn{2}{l}{Ar} &
    \multicolumn{2}{l}{H$_2$O} &
    \multicolumn{2}{l}{O$_2$} &
    \multicolumn{2}{l}{Kr} &
    \multicolumn{2}{l}{Xe} \\

    without $\eta$ &
    8&4  &
    6&2  &
    14&1 &
    9&6  &
    9&0  &
    7&9  &
    9&0  &
    10&7 &
    9&0  &
    9&0  &
    \keV\\

    with $\eta$ &
    3&9 &     
    0&9 &     
    1&1 &
    \multicolumn{2}{c}{} &
    0&33 &
    0&23 &
    4&5 &
    3&1 &
    \multicolumn{2}{c}{} &
    \multicolumn{2}{c}{} &
    \keV\\

    \hline
    \hline

  \end{tabular}

  \caption{Maximum energy at which a stable equilibrium can be
    established in a frictional cooling schemed for \muplus\
    with and without accounting for effective charge for several
    stopping media. A blank entry means no equilibrium energy
    can be established.\label{tab:cooling_ranges}}
\end{table*}


Accounting for the muon's effective charge greatly reduces the maximum
energy at which the condition of
\texteqn~(\ref{eqn:cooling_requirement}) is met.
Table~\ref{tab:cooling_ranges} lists the cooling ranges for several
materials with and without accounting for effective charge.

\begin{figure*}[t]
  \centering
  \ifcolor
    \includegraphics[width=\columnwidth]{\figuresdir 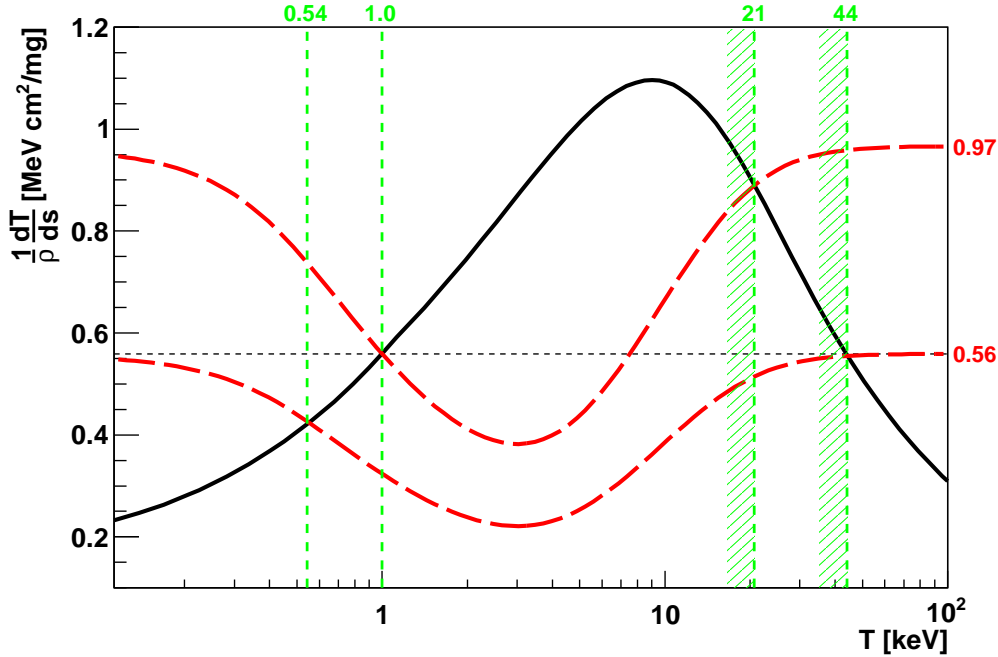}
  \else
    \includegraphics[width=\columnwidth]{\figuresdir dEdx_with_effcharge_He_bw.eps}
  \fi
  \caption{Stopping power of helium on \muplus\ (solid) and
    accelerating power of a uniform constant electric field of
    strength $E$ on \muplus\ in helium with (large dashes\ifcolor,
    red\fi) and without (\ifcolor\else horizontal, \fi small
    dashes\ifcolor, black\fi) accounting for effective charge for two
    values of $eE$ (as indicated on the right
    axis).\label{fig:dedx_with_effcharge_He}}
\end{figure*}

The effects of charge exchange processes also reduce the value of
\TeqPrime, the maximum kinetic energy that can be decelerated to
\Teq. This effect is illustrated in
\textfig~\ref{fig:dedx_with_effcharge_He}, which shows the
frictional-cooling energy region of the stopping power of helium on
\muplus.  Superimposed on the stopping-power curve are three
electric-field accelerating powers: The \ifcolor black\else
horizontal\fi\ dashed line is the naive accelerating power of
\textfig~\ref{fig:dEdx}, with effective charge neglected, for a field
strength \unit{560}{\kilo\Vcmcmmg} resulting in $\Teq=\unit{1}{\keV}$.
The lower \ifcolor red\else dashed\fi\ curve is the accelerating power
for the same electric field strength, accounting for effective charge.
The equilibrium energy is cut in half, but \TeqPrime\ remains the
same. The upper \ifcolor red\else dashed\fi\ curve shows the
accelerating power accounting for effective charge that results in
$\Teq=\unit{1}{\keV}$, which requires a field strength of
\unit{970}{\kilo\Vcmcmmg}, and reduces \TeqPrime\ by a factor of two.

\begin{figure*}[t]
  \centering
  \includegraphics[width=\textwidth]{\figuresdir 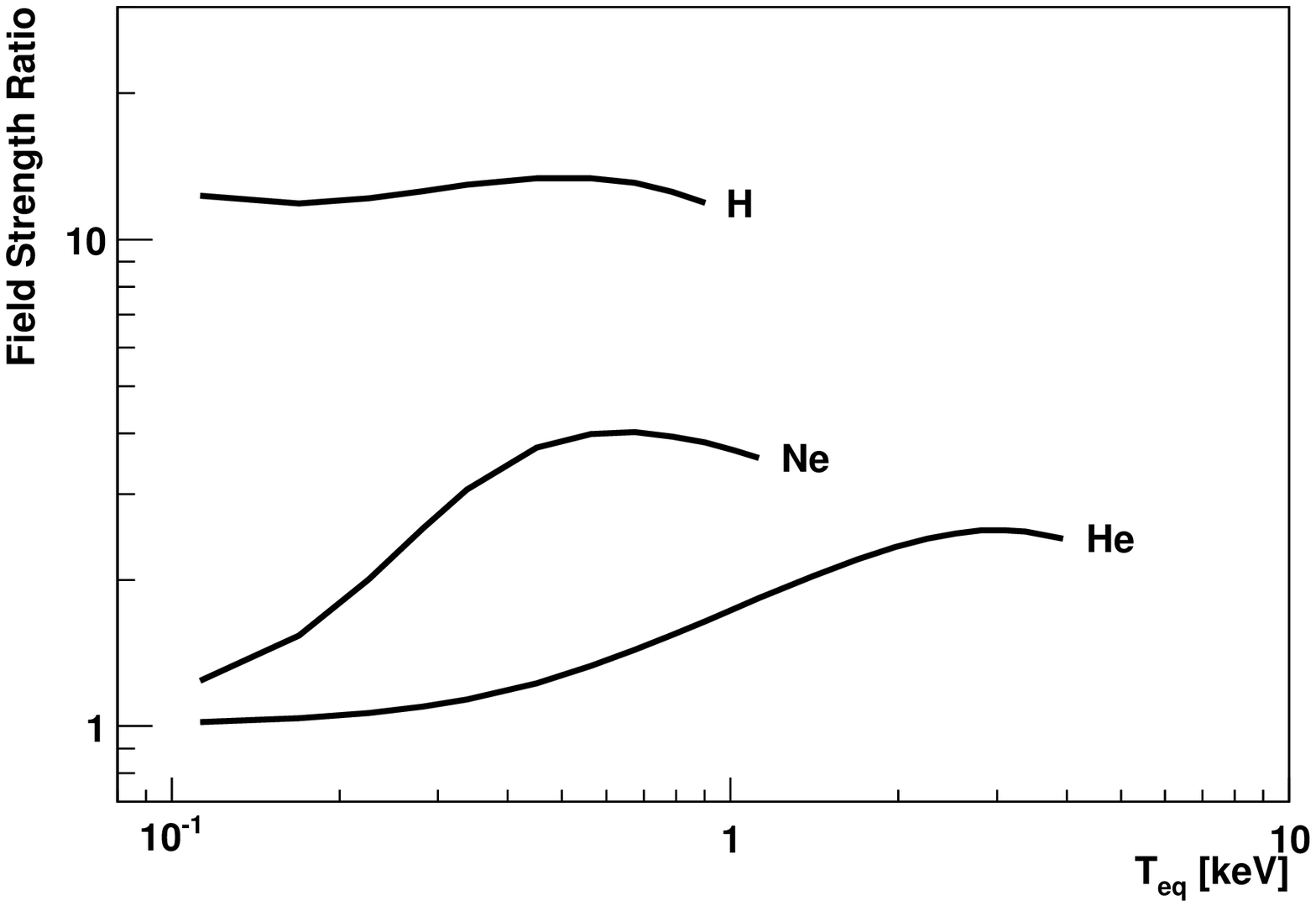}
  \caption{Ratio of electric field strength required to achieve the
    desired equilibrium energy accounting for effective charge to the
    field strength required without accounting for effective
    charge.\label{fig:ap}}
\end{figure*}

Achieving the desired equilibrium energy, accounting for effective
charge, requires a greater electric field strength than is expected in
the naive scheme. The field strength may in fact have to be an order
of magnitude larger, depending on what the desired \Teq\ and the
stopping medium are~(\parfig~\ref{fig:ap}).  On the bright side,
the relative slope of the stopping power is generally larger when
effective charge is accounted for, perhaps causing (according to
\cite{Muhlbauer:1993sb}) the final energy spread of a cooled beam to
be smaller. As well, the stronger electric fields may reduce the
spread caused by scattering.

We performed Monte Carlo simulations of the cooling of \muplus\ beams
with discrete charge exchange interactions, as well as simulations
using the effective-charge approximation. These conformed to the
calculations presented in \textfig~\ref{fig:ap} to within a few
percent.

From the simulations we can calculate the spread of the beam energy
around \Teq\ due solely to charge exchange interactions.
\textTab~\ref{tab:muehlbauer} lists these spreads for \muplus\ in
helium at the same equilibrium energies cooled to
in~\cite{Muhlbauer:1993sb}, which looked at \muminus\ in hydrogen.
The spread due to charge exchange for \muplus\ is smaller than the
spreads due to scattering and straggling for \muminus. A comparison
using \muplus\ in hydrogen is not possible because the maximum \Teq\
in H$_2$ for \muplus\ is \unit{0.9}{\keV}.

The choice of stopping medium is even further limited than the
requirement that the relative slope of the stopping power be positive.
This is illustrated by the example of oxygen, in which the relative
slope of the stopping power is just barely positive over a small
region of energies; but it is not sufficiently large to clearly
establish an equilibrium energy.

Furthermore, at low energies, oxygen's stopping power is proportional
to particle velocity, $S\propto T^{\frac{1}{2}}$, but its $\eta$ is
proportional to $T^{k}$, with $k>\frac{1}{2}$, causing the
accelerating power to decrease below the stopping power at low
energies. A particle that experiences a large loss of energy in one
interaction---or one that scatters into a direction opposite that of
the electric field---will not reaccelerate to \Teq, but rather
continue decelerating to thermal energies and be lost to the cooling
process.  This would severely limit the use of such a medium in a
frictional cooling scheme.

The only viable media for a gaseous frictional cooling scheme for
\muplus\ with equilibrium energies at or above \unit{1}{\keV} are
helium and water vapor. For \Teq\ just below \unit{1}{\keV}, hydrogen
and neon also become viable; and for \Teq\ of a few hundred electron
volts, argon and nitrogen are also viable.

\section{Beam Neutralization \& Foils}

The frictional cooling experiment of \cite{Muh96} iterated energy loss
and replacement by means of a series of moderating foils with electric
fields between them. The experiment and a Monte Carlo
simulation in \cite{Muhlbauer:1993sb} gave promising results for such
a scheme for the cooling of negative muons. In both publications it
was posited that the frictional cooling scheme used with \muminus\
could also work with \muplus; more recently this idea has been
revisited in~\cite{Roberts:2009zzb} and~\cite{Kaplan:2009jd}. However,
the presence of charge exchange interactions greatly limits the yield
for such a scheme; in fact, leading to a zero yield.

A foil-based frictional cooling scheme has the benefit of separating
energy loss, occurring in the foils, from energy restoration,
occurring between foils. Since particles are reaccelerated in vacuum,
the energy restoration depends on the charge state of the particle as
it exits the foil. Those exiting in a neutral state are blind to the
reacceleration field.

The four studies cited above used and simulated carbon foils in their
frictional cooling schemes, choosing the reacceleration field strength
to precisely compensate the energy lost in a foil by a particle at
\Teq; and assumed every particle exits every foil in a charged state.
However, for positively charged particles, after exiting a foil at
mean energy \Teq, a beam will be divided into two populations, a
fraction $f^+(\Teq)$ in the \muplus\ state and a fraction
$(1-f^+(\Teq))$ in the \muonium\ state. The \muplus\ portion of the
beam reaccelerates to \Teq, and upon exiting the next foil in the
cooling series, its population fraction decreases further to
$(f^+(\Teq))^2$. The \muonium\ portion of the beam will not have its
energy losses to the previous foil restored by the electric field and
will exit the next foil at an energy below \Teq, where the chances of
\muonium\ atoms losing their electrons are even smaller.  Furthermore,
\muonium\ atoms that have their electrons stripped off, must exit
several foils in a row in the postive state in order to reaccelerate
back to \Teq. As the beam passes through the array of foils it is
neutralized and slowed down.  Cooling becomes an impractical goal.

\begin{figure*}[t]
  \centering
  \includegraphics[width=\columnwidth]{\figuresdir 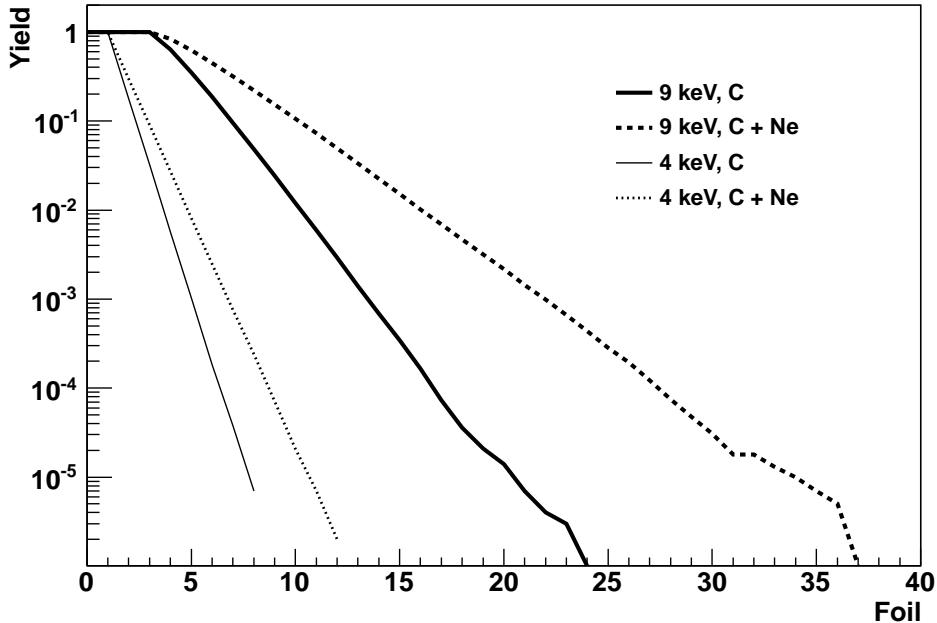}
  \caption{Yield of \muplus\ in carbon-foil frictional cooling schemes
    as a function of number of foils traversed after reaching \Teq,
    for both naked foils and frozen-neon-coated
    foils.\label{fig:foils}}
\end{figure*}

We performed a Monte Carlo simulation of exactly this mechanism. The
foil density was chosen to match the simulations
in~\cite{Muhlbauer:1993sb} and the experiment in~\cite{Muh96},
\unit{5}{\micro\gram\per\centi\meter\squared}~(25-\nano\meter-thick
foils of 2-\gram\per\centi\meter\cubed\ amorphous carbon).  Scattering
and energy-loss straggling were neglected, since they would only
worsen the beam neutralization described above.
\textFig~\ref{fig:foils} shows the yield\footnote{In the simulation,
  tracking of a particle is terminated when it does not have
  sufficient kinetic energy to pass completely through a single foil
  according to~\cite{icru49:1994tf}. The yield is thus defined as the
  fraction of muons surviving passage through the foil.} of the
foil-based scheme as a function of the number of foils traversed for
mono-energetic beams starting at \unit{9}{\keV}, the maximum \Teq\
possible, and \unit{4}{\keV}, the \Teq\ used
in~\cite{Muhlbauer:1993sb}. After a handful of foils, the yield falls
off exponentially with the number of foils traversed.  After passing
through the first foil and energy restoration, the mean energy of the
beam is greatly reduced from \Teq\ and the energy spread is greatly
enlarged. The mean energy continues to fall with each successive foil.

We also simulated a scheme in which the foils are coated with frozen
noble gas to increase the fraction of particles exiting them in the
positive charge state. This is inspired by the scheme for the
production of low-energy muons from a surface muon beam at the Paul
Scherrer Institute (\PSI), which uses frozen argon and
krypton~\cite{PhysRevLett.72.2793}. Neither argon nor krypton coatings
improve the beam neutralization. A neon coating marginally slows beam
neutralization (shown in \parfig~\ref{fig:foils}), but does not
prevent the degradation of the beam energy. Moreover, we simulated the
charge-exchange effects of the coatings but neglected their impact on
energy loss. This assumes that the coatings are nanometers thin;
however, the coatings used at \PSI\ are a much thicker
\unit{200-300}{\nano\meter}. Such thicker coatings would hasten beam
loss.

The scheme of~\cite{Roberts:2009zzb}, also based on carbon (graphite)
foils, requires the beam to pass through order-100 foils at the
equilibrium energy. A beam of positively charged muons (or protons)
would be unable to survive such a scheme.

\section{Conclusion}

Muon collider schemes employing frictional cooling are a viable option
for collision of multi-\TeV\ lepton beams. Several articles have been
published with analytical and experimental results for frictional
cooling of negatively charged particles. Many of these articles have
conjectured that the results for positively charged particles will be
the same as for negatively charged ones, and schemes for the cooling
of positive particles have been proposed. A key group of physics
processes involved in the slowing down of positive particles---those
changing the charge state---has been neglected in these studies. We
found that accounting for these processes significantly alters the
results for positive particles from those for negative ones: The
choice of cooling medium is greatly limited, such that helium gas
becomes the only viable medium; with foil-based schemes completely
ruled out. The range of equilibrium energies for the cooled beam is
also greatly limited, with a maximum possible energy of approximately
\unit{4}{\keV} for \muplus~(\unit{36}{\keV} for protons). And the
electric field strength required to bring a beam of positive particles
to an equilibrium energy is significantly greater than the strength
required to bring a beam of negative ones to the same energy.

\ifarxiv

\else
  \section*{References}
  \bibliographystyle{bst/h-physrev}
  \bibliography{effcharge_paper}
\fi

\end{document}